\documentclass[aps, prb, twocolumn, showpacs]{revtex4}
\usepackage{graphicx}
\usepackage{bm}

\begin{document}

\title{Quantum phase transitions of the diluted O(3) rotor model}

\author{Thomas Vojta}
\affiliation{Department of Physics, University of Missouri - Rolla, Rolla, MO 65409, USA}

\author{Rastko Sknepnek}
\affiliation{Department of Physics and Astronomy, McMaster University, Hamilton ON L8S
4M1, Canada}

\date{\today}

\begin{abstract}
We study the phase diagram and the quantum phase transitions of a site-diluted
two-dimensional O(3) quantum rotor model by means of large-scale Monte-Carlo simulations.
This system has two quantum phase transitions, a generic one for small dilutions, and a
percolation transition across the lattice percolation threshold. We determine the
critical behavior for both transitions and for the multicritical point that separates
them. In contrast to the exotic scaling scenarios found in other random quantum systems,
all these transitions are characterized by finite-disorder fixed points with power-law
scaling. We relate our findings to a recent classification of phase transitions with
quenched disorder according to the rare region dimensionality, and we discuss experiments
in disordered quantum magnets.
\end{abstract}


\maketitle

\section{Introduction}
\label{sec:intro}

Quantum phase transitions occur at zero temperature when a nonthermal parameter like
pressure, magnetic field, or chemical composition is varied. In the presence of defects,
impurities, and other kinds of quenched disorder, the interplay between dynamic quantum
fluctuations and static disorder fluctuations can lead to a variety of unconventional
phenomena. Experimental examples include quantum Ising spin glasses,
\cite{WERAR91,WBRA93} heavy-fermion intermetallic compounds,
\cite{SMLG91,AndrakaTsvelik91,ACDD98,Stewart01} and other itinerant quantum magnets
\cite{DGYM03} as well as high-temperature superconductors,
\cite{AttfieldKharlanovMcAllister98,PTRX02} the metal-insulator transition in
metal-oxide-semiconductor field effect transistors (MOSFETs),
\cite{AbrahamsKravchenkoSarachik01,KMBF95} and superconductor-insulator transitions in
thin films. \cite{HebardPaalanen90}

Quenched disorder has interesting consequences already at classical phase transitions. In
the early years, it was thought that impurities always destroy a critical point, because
the system divides itself up into spatial regions that undergo the transition at
different temperatures (see discussion in Ref.\ \onlinecite{Grinstein85} and references
therein). However, it soon became clear that in classical systems with short-range
disorder correlations, the transition generically remains sharp. Harris \cite{Harris74}
derived a criterion for the stability of a clean critical point against disorder: If the
correlation length exponent $\nu$ fulfills the inequality $\nu>2/d$, where $d$ is the
spatial dimensionality, the critical behavior is not influenced by weak disorder. Harris'
idea can be generalized to form the basis of a classification of critical points
according to the behavior of the disorder strength under coarse graining. \cite{MMHF00}
The first class contains systems that fulfill the Harris criterion. In these systems, the
disorder decreases without limit under coarse graining. The critical behavior is governed
by a clean renormalization group fixed point, and macroscopic observables are
self-averaging. The other two classes can occur when the clean system violates the Harris
criterion. In systems belonging to the second class, the disorder strength approaches a
nonzero constant for large length scales, corresponding to a fixed point with finite
disorder. The critical exponent are thus different from that of the corresponding clean
system, with the new dirty correlation length exponent fulfilling the inequality
$\nu>2/d$.\cite{CCFS86} Moreover, macroscopic observables are not self-averaging.
\cite{AharonyHarris96,WisemanDomany98} Finally, the third class contains systems in which
the disorder strength (counter-intuitively) \emph{increases} without limit under coarse
graining. The resulting infinite-randomness fixed point has unconventional properties
including exponential rather than power-law scaling and very broad distributions of
macroscopic observables. \cite{Fisher92,Fisher95}

At zero-temperature quantum phase transitions, order-parameter fluctuations in space and
time must be considered. \cite{Hertz76,Sachdev_book99} Quenched disorder is
time-independent, it is thus perfectly correlated in one of the relevant dimensions, the
(imaginary) time dimension. Because these correlations increase the effects of the
disorder, quantum phase transitions are generically more strongly affected by disorder
than classical transitions, potentially resulting in unconventional behavior. One of the
earliest explicit examples was the random transverse-field Ising chain
\cite{Fisher92,Fisher95,YoungRieger96} (or the equivalent McCoy-Wu model
\cite{McCoyWu68,McCoy69}). This system belongs to the third of the classes discussed
above, i.e., the critical point is of infinite-randomness type. The dynamical scaling is
activated with the correlation time $\xi_\tau$ and correlation length $\xi$ being related
by $\ln \xi_\tau\sim \xi^\psi$. (In contrast, at conventional critical points, this
relation is a power law, $\xi_\tau\sim \xi^z$, with a universal dynamical exponent $z$).
Analogous behavior has been found, e.g., in the two-dimensional transverse-field Ising
model \cite{PYRK98,MMHF00} and in quantum Ising spin glasses.
\cite{ThillHuse95,RiegerYoung96}

An important aspect of phase transitions in disordered systems are the so-called rare
regions, large spatial regions that are devoid of impurities or more strongly coupled
than the bulk system. These regions can be in the ordered phase even though the
bulk system is still in the disordered phase. Griffiths \cite{Griffiths69} showed that
this leads to a singularity (the Griffiths singularity) in the free energy in an entire
parameter region (the Griffiths region or Griffiths phase \cite{RanderiaSethnaPalmer85})
close to the phase transition. In generic classical systems with short-range disorder
correlations, thermodynamic Griffiths effects are very weak because the singularity in
the free energy is only an essential one. They are therefore probably unobservable in
experiment. However, disorder correlations can greatly enhance the rare region effects.
\cite{McCoyWu68,McCoy69}

Since quenched disorder is perfectly correlated in the (imaginary) time direction,
quantum phase transitions are expected to display stronger rare region effects than
classical transitions. Indeed, in the above-mentioned random quantum Ising systems, the
Griffiths singularities are of power-law type with the susceptibility diverging over a
finite parameter range.\cite{Fisher92,Fisher95,YoungRieger96,ThillHuse95,RiegerYoung96}
In itinerant quantum magnets, rare region effects can be even more dramatic. For Ising
symmetry, the sharp quantum phase transition is destroyed by smearing \cite{Vojta03a}
because sufficiently large rare regions stop tunneling. The same also happens in
classical Ising magnets with plane defects\cite{Vojta03b,SknepnekVojta04} and at certain
nonequilibrium phase transitions. \cite{Vojta04,DickisonVojta05} A recent review of these
and other rare region effects can be found in Ref.\ \onlinecite{Vojta06}.

In systems with continuous order parameter symmetry, the situation is more complex. The
ground states of certain one-dimensional quantum spin chains are controlled by
infinite-randomness fixed points. \cite{Fisher94} On the other hand, in dimensions $d \ge
2$, the {\em stable low-energy} fixed point of random Heisenberg models has been shown to
be conventional. \cite{LMRI03,LWLR06} Preliminary renormalization group results
\cite{MMHF00} for the critical point in these models suggested that the
infinite-randomness fixed point is unstable, implying more conventional behavior. This
agrees with Monte-Carlo simulations of diluted single-layer \cite{Sandvik01,Sandvik02b}
or bilayer \cite{Sandvik02,VajkGreven02} quantum Heisenberg antiferromagnets that did not
show indications of exotic scaling. Note, however, that inhomogeneous bond disorder can
induce a new quantum-disordered phase that can be understood as a quantum Griffiths
phase. \cite{YuRoscildeHaas05,YuRoscildeHaas06}

In this paper, we report the results of large-scale Monte-Carlo simulations of a diluted
O(3) quantum rotor model in two space dimensions. We find that the system has two quantum
phase transitions, a generic one for dilutions below the lattice percolation threshold
$p_c$ and a percolation type transition right at $p_c$. Both transitions and the
multicritical point that separates them  display conventional power-law critical
behavior. For the generic transition, the critical exponents are universal, i.e.,
independent of the dilution. A short account of part of this work has already been
published in Ref.\ \onlinecite{SknepnekVojtaVojta04}. The present paper is organized as
follows: In section \ref{sec:theory}, we introduce the model and summarize the scaling
theories for conventional and infinite-randomness critical points. The simulation method
and our results for the phase diagram and the critical behavior of the quantum phase
transitions are presented in section \ref{sec:simulations}. In the concluding section
\ref{sec:conclusions} we relate our results to a general classification
\cite{VojtaSchmalian05} of dirty phase transitions, and we consider experiments.

\section{Theory}
\label{sec:theory}
\subsection{Diluted quantum rotor model}
\label{subsec:model}

We consider a site-diluted O(N) quantum rotor model defined on a square lattice. Its
quantum Hamiltonian is given by \cite{Sachdev_book99}
\begin{equation}
\hat H_Q = U \sum_i \epsilon_i \, \hat \mathbf{L}_i^2 - J \sum_{\langle i,j \rangle}
\epsilon_i \epsilon_j \, \hat\mathbf{n}_i \cdot \hat\mathbf{n}_j ~. \label{eq:H_Q}
\end{equation}
Here, $\hat\mathbf{n}_i$ is an $N$-component unit vector at site $i$. Conjugate momenta
$\hat\mathbf{p}_i$ are defined via the usual canonical commutation relations
$[\hat{n}^\alpha,\hat{p}^\beta]=i\delta_{\alpha\beta}$ on each site $i$.
($\alpha,\beta=1\ldots N$ are the component indices, and we work in units in which
$\hbar=1$.) The components of the angular momentum $\hat \mathbf{L}$ of each rotor are
given by $\hat L^{\alpha\beta} = \hat{n}^\alpha \hat{p}^\beta - \hat{n}^\beta
\hat{p}^\alpha$. The site dilution is described by the independent random variable
$\epsilon_i$ which can take the value 0 and 1 with probability $p$ and $1-p$,
respectively.

\begin{figure}
\centerline{
\includegraphics[width=0.85\columnwidth]{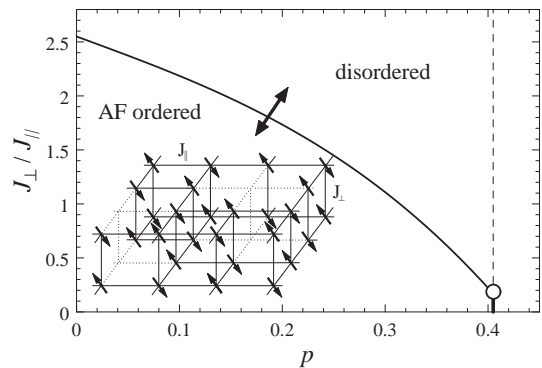}
} \caption{ Phase diagram of the diluted bilayer Heisenberg antiferromagnet, as function
of $J_\perp/J_\parallel$ and dilution $p$. The dashed line is the percolation threshold,
the open dot is the multicritical point of Refs.~\onlinecite{Sandvik02, VajkGreven02}.
The arrow indicates the QPT studied here. Inset: The model: Quantum spins (arrows)
$\mathbf{\hat{S}}_{i,1}$ and $\mathbf{\hat{S}}_{i,2}$ reside on the two parallel square
lattices. The spins in each plane interact with the coupling strength $J_\parallel$.
Interplane coupling is $J_\perp$. Dilution is done by removing dimers.} \label{fig:model}
\end{figure}

Elementary quantum rotors do not exist in nature; rather, they arise as effective
low-energy degrees of freedom of correlated quantum systems. For example, O(2) quantum
rotor models describe superconducting Josephson junction arrays or bosons in optical
lattices. An $N$=3 quantum rotor describes the states of an even number of
antiferromagnetically coupled Heisenberg spins. A specific example is provided by the
bilayer quantum Heisenberg antiferromagnet depicted in the inset of Fig.\
\ref{fig:model}.
This system is equivalent to an O(3) quantum rotor model with each dimer
$(\mathbf{\hat{S}}_{i,1},\mathbf{\hat{S}}_{i,2})$ of spins at site $i$ and layers 1 and 2
being represented by a single rotor. The rotor coordinate $\mathbf{\hat{n}}_i$
corresponds to $\mathbf{\hat{S}}_{i,1} - \mathbf{\hat{S}}_{i,2}$ and the angular momentum
$\mathbf{\hat{L}}_i$ corresponds to $\mathbf{\hat{S}}_{i,1} + \mathbf{\hat{S}}_{i,2}$
(see, e.g., chapter 5 of Ref. \onlinecite{Sachdev_book99}). The O(3) quantum rotors also
describe double-layer quantum Hall ferromagnets.

\begin{figure}
\centerline{
\includegraphics[width=0.7\columnwidth]{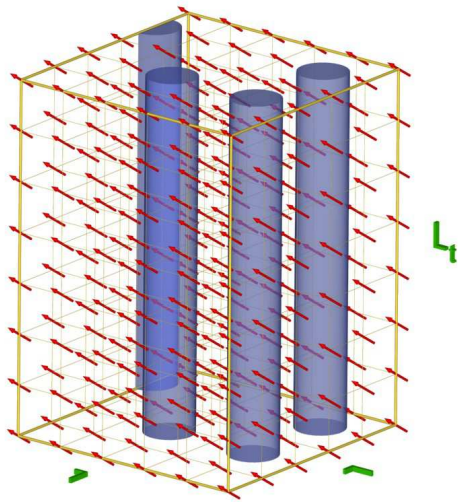}
} \caption{Sketch of the classical model (\ref{eq:H}). The arrows represent the classical
spins and the tubes show the locations of the linear defects (vacancies).}
\label{fig:swisscheese}
\end{figure}

We now focus on the O(3) site-diluted quantum rotor model on a square lattice. Since we
will be mostly interested in the universal critical behavior, we map the quantum system
onto a classical system in the same universality class. This can be easily achieved via a
path integral representation of the partition function. \cite{Sachdev_book99} The
resulting classical system is a three-dimensional Heisenberg model with the extra
dimension representing the imaginary time coordinate of the quantum rotor model. Because
the impurities in the quantum rotor model are quenched (i.e., time-independent), the
defects in the classical system are linear, i.e., the disorder is perfectly correlated in
the extra (imaginary time) direction (see Fig.\ \ref{fig:swisscheese}).
Thus, our classical Hamiltonian reads:
\begin{equation}
\label{eq:H} H=K\sum_{\langle i,j\rangle, \tau}
\epsilon_i\epsilon_j\mathbf{n}_{i,\tau}\cdot\mathbf{n}_{j,\tau}+K\sum_{i,\tau}\epsilon_i\mathbf{n}_{i,\tau}\cdot\mathbf{n}_{i,\tau+1},
\end{equation}
where $\mathbf{n}_{i,\tau}$ is an O(3) unit vector at the lattice site with spatial
coordinate $i$ and ``imaginary time'' coordinate $\tau$. The coupling constant $\beta K$
of the classical model is related to the ratio $U/J$ of the quantum rotor model. Here,
$\beta\equiv1/T$ where $T$ is an effective ``classical'' temperature, not equal to the
real temperature in the quantum model which is zero. We set $K=1$ and drive the classical
system through the transition by tuning the classical temperature $T$.

As an aside, we note that in the above-mentioned bilayer Heisenberg antiferromagnet the
dilution has to be done by removing dimers of corresponding spins in the two layers
because each dimer is described by a single rotor. In contrast, for site dilution, the
physics changes completely: Random Berry phase terms with no classical analogue arise.
They are equivalent to impurity-induced moments, \cite{SachdevVojta01} and those become
weakly coupled via bulk excitations. Thus, for all dilutions below the percolation
threshold, $p<p_c$, the ground state shows long-range order, independent of the coupling
constants! This effect is absent for dimer dilution, and both phases of the clean system
survive for small dilution.

The critical behavior of the Hamiltonian (\ref{eq:H}) in the clean limit ($p=0$) is in
the three-dimensional classical Heisenberg universality class, with the correlation
length exponent $\nu \approx0.7$. It thus violates the Harris criterion $\nu >
2/d_\bot=1$ ($d_\bot=2$ because only dimensions in which there is randomness count for
the Harris criterion). Therefore, the critical behavior must change upon diluting the
lattice.

\subsection{Power-law vs. activated scaling}
\label{subsec:scaling}

In this subsection, we summarize the conventional and activated scaling scenarios at
critical points with quenched disorder to the extent necessary for the analysis of our
simulation results.

At conventional (quantum) critical points, correlation length $\xi$ and correlation time
$\xi_\tau$ are related by a power-law, $\xi_\tau \propto \xi^z$ with $z$ being the
dynamical critical exponent. (Note that in the effective classical system (\ref{eq:H}),
$\xi$ and $\xi_\tau$ are the correlation lengths in the space-like and time-like
directions, respectively.) This is referred to as power-law dynamical scaling. In
contrast, at infinite-randomness critical points, the dynamical scaling is activated,
i.e., the relation between correlation length and time is exponential, $\ln (\xi_\tau)
\propto \xi^\psi$.\cite{Fisher92,Fisher95}

These differences in the dynamical scaling lead to analogous differences in the
finite-size scaling behavior of observables. If we denote the linear system size in the
two space-like dimensions by $L$ and the size in the time-like dimension by $L_\tau$, the
finite-size scaling forms of the magnetization per site $m=|\mathbf{m}|$ and the
susceptibility $\chi$ at a {conventional} critical point read
\begin{eqnarray}
m &=& L^{-\beta/\nu} \tilde m_C(tL^{1/\nu},L_\tau/L^z)~, \label{eq:m_conv}\\ \chi &=&
L^{\gamma/\nu} \tilde \chi_C(tL^{1/\nu},L_\tau/L^z)~. \label{eq:chi_conv}
\end{eqnarray}
Here, $t$ is the dimensionless distance from the critical point; and $\beta,\gamma$ and
$\nu$ are the critical exponents of magnetization, susceptibility, and correlation
length, respectively. At an infinite-randomness critical point, the scaling combination
$L_\tau/L^z$ has to be replaced by $\ln(L_\tau)/L^\psi$ leading to the finite-size
scaling forms
\begin{eqnarray}
m &=& L^{-\beta/\nu} \tilde m_A(tL^{1/\nu},\ln(L_\tau)/L^\psi)~, \\
\chi &=& L^{\gamma/\nu} \tilde \chi_A(tL^{1/\nu},\ln(L_\tau)/L^\psi)~.
\label{eq:mchi_act}
\end{eqnarray}

In addition to magnetization and susceptibility we also calculate three quantities whose
scale dimension is zero which makes them particularly suitable for locating the critical
point and extracting high precision values for the correlation length and dynamical
exponents. The first such quantity is the Binder ratio. It is defined by
\begin{equation}
\label{eq:Binder} g_{av}=\left[ 1-\frac{\langle |\mathbf{m}|^4\rangle}{3\langle
|\mathbf{m}|^2\rangle^2}\right]_{av},
\end{equation}
where $\left[\ldots\right]_{av}$ denotes the disorder average and $\langle\ldots\rangle$
denotes the Monte-Carlo average for each sample. This quantity approaches well-known
limits in both bulk phases (stable fixed points): In the ordered phase, all spins are
correlated, and the magnetization has small fluctuations around a nonzero value.
Therefore, $\langle|\mathbf m|^4 \rangle \approx \langle|\mathbf m|^2\rangle^2$, and the
Binder ratio approaches 2/3. In the disordered phase, the system consists of many
independent fluctuators. Consequently, $\langle|\mathbf m|^4 \rangle$ can be decomposed
using Wick's theorem. For O(3) symmetry this gives $\langle|\mathbf m|^4 \rangle \approx
(15/9) \langle|\mathbf m|^2\rangle^2$, and the Binder ratio approaches 4/9. More
generally, the Binder ratio is large if all spins are correlated and decreases if the
system contains independently fluctuation units. Because the Binder ratio has scale
dimension zero, its finite-size scaling form is given by
\begin{eqnarray}
\label{eq:g_conv} g_{av}&=&\tilde{g}_C (tL^{1/\nu},L_\tau/L^z)\qquad {\rm or}
\\ \label{eq:g_act} g_{av}&=&\tilde{g}_A (tL^{1/\nu},\ln(L_\tau)/L^\psi)
\end{eqnarray}
for conventional scaling or for activated scaling, respectively. Two important
characteristics follow from the scaling form and the discussion
above:\cite{GuoBhattHuse94,RiegerYoung94} (i) For fixed $L$, $g_{av}$ has a peak as a
function of $L_{\tau}$. The peak position $L_{\tau}^{\rm max}$ marks the {\em optimal}
sample shape, where the ratio $L_{\tau}/L$ roughly behaves like the corresponding ratio
of the correlation lengths in time and space directions, $\xi_{\tau}/\xi$. (If the aspect
ratio deviates from the optimal one, the system can be decomposed into independent units
either in space or in time direction, and thus $g_{av}$ decreases.) At the critical
temperature $T_c$, the peak value $g_{av}^{\rm max}$ is independent of $L$. Thus, for
power law scaling, plotting $g_{av}$ vs. $L_\tau/L_\tau^{max}$ at $T_c$ should collapse
the data, without the need for a value of $z$. In contrast, for activated scaling the
$g_{av}$ data should collapse when plotted as a function of
$\log(L_\tau)/\log(L_\tau^{\rm max})$. (ii) For samples of the optimal shape
($L_\tau=L_\tau^{max}$), plots of $g_{av}$ vs. temperature for different $L$ cross at
$T_c$.

The other two quantities of scale dimension zero we consider, are the ratios of
disconnected correlation lengths and system sizes in both space and time-like dimensions.
Here, the disconnected correlation lengths $\xi^{dis}$ and $\xi_\tau^{dis}$ arise from
the disconnected correlation function $\langle
\mathbf{n}_{i,\tau}\mathbf{n}_{j,\tau^\prime} \rangle$. In contrast the usual, connected
correlation lengths $\xi$ and $\xi_\tau$ arise from the connected correlation function
$\langle \mathbf{n}_{i,\tau}\mathbf{n}_{j,\tau^\prime} \rangle -\langle
\mathbf{n}_{i,\tau}\rangle\langle\mathbf{n}_{j,\tau^\prime} \rangle$. The finite-size
scaling forms of our ratios for conventional and activated scaling read
\begin{eqnarray}
\label{eq:xiL_conv} \xi^{dis}/L&=&X_C (tL^{1/\nu},L_\tau/L^z)\qquad {\rm or}
\\ \label{eq:xiL_act} \xi^{dis}/L&=&X_A (tL^{1/\nu},\ln(L_\tau)/L^\psi)~,
\end{eqnarray}
and
\begin{eqnarray}
\label{eq:xitL_conv} \xi_\tau^{dis}/L_\tau&=&Y_C (tL^{1/\nu},L_\tau/L^z)\qquad {\rm or}
\\ \label{eq:xitL_act} \xi_\tau^{dis}/L_\tau&=&Y_A (tL^{1/\nu},\ln(L_\tau)/L^\psi)~,
\end{eqnarray}
respectively. Calculating these quantities provides independent checks for the location
of the critical point and for the exponents $z$ (or $\psi$) and $\nu$.

\section{Simulations}
\label{sec:simulations}
\subsection{Monte Carlo method}
\label{subsec:method}

In order to study the phase transitions of the effective classical model (\ref{eq:H}) we
perform large scale Monte-Carlo simulations. We use the efficient Wolff cluster algorithm
\cite{Wolff89} to reduce the effects of the critical slowing down close to the phase
transition. This is possible because the dilution-disorder does not introduce
frustration, and all interactions are ferromagnetic.  We investigate linear sizes up to
$L=120$ in space direction and $L_\tau=2560$ in imaginary time direction, for impurity
concentrations $p=\frac{1}{8}$, $\frac 1 5$, $\frac 2 7$, $\frac 1 3$ and $p_c=0.407253$
which is the lattice percolation threshold. For the larger dilutions, $p=\frac 2 7, \frac
1 3$, and $p_c$ we perform both Wolff and Metropolis sweeps to equilibrate small dangling
clusters.

The determination of averages, variances, and distribution functions of observables in
disordered systems requires the simulation of many independent samples with different
impurity configurations. Because of the huge computational effort involved,\cite{SelkeShchurTapalov94}
 one must carefully choose the number $N_S$ of disorder
realizations (i.e., samples) and the number $N_I$ of measurements during the simulation
of each sample for optimal performance. Assuming full statistical independence between
different measurements (quite possible with a cluster update), the variance $\sigma_T^2$
of the final result (thermodynamically and disorder averaged) for a particular observable
is given by\cite{BFMM98,BFMM98b}
\begin{equation}
\sigma_T^2 = (\sigma_S^2 + \sigma_I^2/N_I)/N_S
\end{equation}
where $\sigma_S$ is the disorder-induced variance between samples and $\sigma_I$ is the
variance of measurements within each sample. Since the computational effort is roughly
proportional to $N_I N_S$ (neglecting equilibration for the moment), it is then clear
that the optimum value of $N_I$ is very small. One might even be tempted to measure only
once per sample. On the other hand, with too short measurement runs most computer time
would be spent on equilibration.

In order to balance these requirements we have used a large number $N_S$ of disorder
realizations, ranging from 1000 to several 10000, depending on the system size and rather
short runs of 100-200 Monte-Carlo sweeps, with measurements taken after every sweep. (A
sweep is defined by a number of cluster flips so that the total number of flipped spins
is equal to the number of sites, i.e., on the average each spin is flipped once per
sweep.) The length of the equilibration period for each sample is also 100 Monte-Carlo
sweeps. The actual equilibration times have typically been of the order of $10$-$20$
sweeps at maximum. Thus, an equilibration period of 100 sweeps should be more than
sufficient.

\subsection{Results: phase diagram}
\label{subsec:pd}

We start the discussion of our results by considering the phase diagram of the classical
Hamiltonian (\ref{eq:H}) in the dilution-temperature plane. To determine the critical
temperature $T_c$ for a given dilution $p$, we use a simple iterative procedure based on
the properties of the Binder ratio $g_{av}$ discussed after (\ref{eq:g_act}). We start
with a guess for the dynamical exponent $z$ (or, alternatively $\psi$ for activated
scaling). We then perform a number of simulation runs to calculate $g_{av}$ as a function
of temperature for samples whose linear sizes fulfill $L_\tau  \propto L^z$. The
approximate crossing of the $g_{av}$ vs. $T$ curves for different $L$ gives an estimate
for $T_c$. At this temperature, we now calculate $g_{av}$ as a function of $L_\tau$ for
fixed $L$. The maxima of these curves give an improved estimate for the ``optimal
shapes'', i.e., for the dynamical exponent $z$. This procedure can be repeated until the
estimate for $T_c$ converges. Typically, only two to three iterations were necessary for
the desired accuracy.

\begin{figure}
\centerline{
\includegraphics[width=0.85\columnwidth]{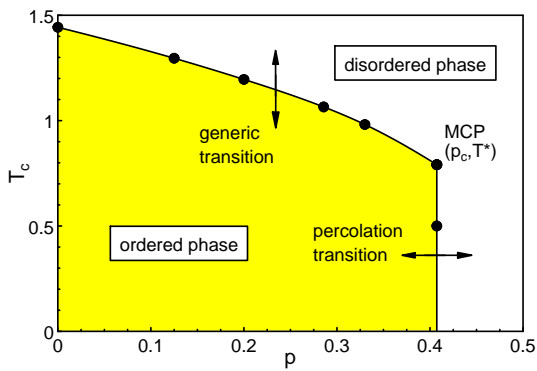}
} \caption{Phase diagram of the three-dimensional Heisenberg model (\ref{eq:H}) as
function of temperature $T$ and concentration $p$ of linear defects. MCP is the
multicritical point. The big dots mark the numerically determined transition points. The
lines are guides for the eye.} \label{fig:pd_classical}
\end{figure}

The resulting phase diagram is shown in Fig.\ \ref{fig:pd_classical}.
As expected, $T_c(p)$ decreases with increasing dilution $p$, but the ordered phase
survives up to the lattice percolation threshold $p_c$ with $T^\ast=T_c(p_c)>0$. Thus,
the classical Heisenberg model (\ref{eq:H}) with linear defects has two phase
transitions, viz., a generic transition for $p<p_c$ and a percolation type transition for
$p=p_c$ and $T<T^\ast$. They are separated by a multicritical point at $(p_c,T^\ast)$.
Analogous behavior was found in the dimer-diluted bilayer quantum  Heisenberg
antiferromagnet\cite{Sandvik02,VajkGreven02} (see also Fig.\ \ref{fig:model}).

We have carried out detailed investigations of both transitions and of the multicritical
point. Our results for the critical behaviors will be presented in the next three
subsections.

\subsection{Generic critical point for $p<p_c$}
\label{subsec:generic}

To determine the critical behavior of the generic transitions and to test its
universality, we have considered four different impurity concentrations, $p=\frac{1}{8}$,
$\frac 1 5$, $\frac 2 7$, and $\frac 1 3$. For each concentration we have performed two
types of simulations: The first consists of runs right at $T_c(p)$ for systems of
different sizes $L$ and $L_\tau$ with varying aspect ratio $L_\tau/L$. The finite-size
scaling properties of $g_{av}$, $m$, and $\chi$ allow us to extract the dynamical
exponent $z$, as well as $\beta/\nu$ and $\gamma/\nu$. In the second set of simulations,
we vary the temperature over a range in the vicinity of $T_c$, but we consider only
samples of ``optimal shape'', $L_\tau \propto L^z$, using the value of $z$ found in the
first part. Finite-size scaling then yields the correlation length exponent $\nu$.

\begin{figure}
\includegraphics[width=\columnwidth]{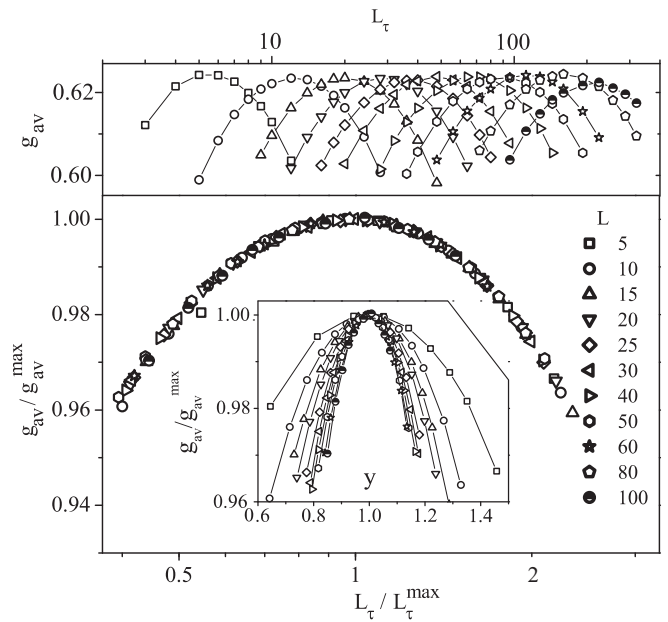}
\caption{ Upper panel: Binder ratio $g_{av}$ as a function of $L_\tau$ for various $L$
($p=\frac 1 5$). Lower panel: Power-law scaling plot $g_{av}/g_{av}^{max}$ vs.
$L_\tau/L_\tau^{max}$  Inset: Activated scaling plot $g_{av}/g_{av}^{max}$ vs.
$y=\log(L_\tau)/\log(L_\tau^{max})$. The statistical errors of the data in the two main
panels are smaller than the symbol size (see text).} \label{fig:zscaling}
\end{figure}

Let us start by discussing the behavior of the Binder ratio $g_{av}$ right at the
critical temperature $T_c(p)$.  The upper panel of Fig.\ \ref{fig:zscaling} shows
$g_{av}$ as a function of $L_\tau$ for various $L=5$ to 100 and dilution $p=\frac 1 5$ at
$T=T_c=1.1955$.
The statistical error of $g_{av}$ is below $0.1\%$ for the smaller sizes and not more
than $0.2\%$ for the largest systems. As expected at $T_c$, the maximum Binder ratio for
each of the curves does not depend on $L$. We now discriminate between power-law
dynamical scaling (\ref{eq:g_conv}) and activated dynamical scaling (\ref{eq:g_act}). To
this end, we plot $g_{av}/g_{av}^{max}$ as a function of $L_\tau/L_\tau^{max}$ in the
lower panel of Fig.\ \ref{fig:zscaling}. The data scale extremely well, giving
statistical errors of $L_\tau^{\rm max}$ in the range between $0.3\%$ and $1\%$.  For
comparison, the inset shows a plot of $g_{av}$ as a function of
$\log(L_\tau)/\log(L_\tau^{max})$ corresponding to activated dynamical scaling
(\ref{eq:g_act}). Plotted this way, the data clearly do not scale. The results for the
other impurity concentrations $p=\frac 1 8, \frac 2 7, \frac 1 3$ are completely
analogous.

\begin{figure}
\includegraphics[width=0.85\columnwidth]{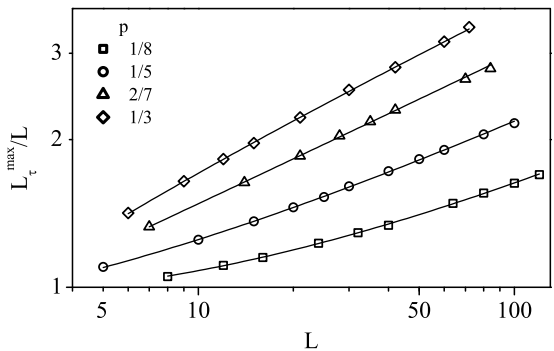}
\caption{ $L_\tau^{max}/L$ vs. $L$ for dilutions $p = \frac 1 8$, $\frac 1 5$, $\frac 2
7$ and $\frac 1 3$. Solid lines: Fit to $L_\tau^{max}=aL^z(1+bL^{-\omega_1})$ with
$z=1.310(6)$ and $\omega_1=0.48(3)$. The statistical errors of the data are well below a
symbol size. } \label{fig:zfit}
\end{figure}

This analysis establishes that the dynamical scaling at the generic transition is of
conventional power-law type. We now proceed to determine the dynamical exponent $z$ and
to study whether or not it is universal, i.e., independent of the dilution $p$. According
to (\ref{eq:g_conv}), the maximum positions $L_\tau^{max}$ should depend on $L$ via a
power law with the exponent $z$. In Fig.\ \ref{fig:zfit}, we plot $L_\tau^{max}$ vs. $L$
for all four dilutions $p$.
The curves show significant deviations from pure power-law behavior which can be
attributed to corrections to scaling due to irrelevant operators. In such a situation, a
direct power-law fit of the data will only yield {\em effective} exponents. To find the
true {\em asymptotic} exponents we take the leading correction to scaling into account by
using the ansatz $L_\tau^{max}(L)=aL^z(1+bL^{-\omega_1})$  with universal
(dilution-independent) exponents $z$ and $\omega_1$ but dilution-dependent $a$ and $b$. A
combined fit of all four curves gives $z=1.310(6)$ and $\omega_1=0.48(3)$ where the
number in brackets is the \emph{statistical} error of the last given digit. The fit is of
high quality ($\chi^2\approx 0.7$) and robust against removing complete data sets or
removing points form the lower or upper end of each set. We thus conclude that the
asymptotic dynamical exponent $z$ is indeed universal. Note that the leading corrections
to scaling vanish very close to $p=\frac 2 7$; the curvature of the $L_\tau^{\rm max}(L)$
curves in Fig.\ \ref{fig:zfit} is opposite above and below this concentration. A straight
power-law fit of the $L_\tau^{max}$ vs.\ $L$ curve for this dilution gives $z=1.303(3)$
in good agreement with the value from the global fit.

\begin{figure}
\includegraphics[width=0.85\columnwidth]{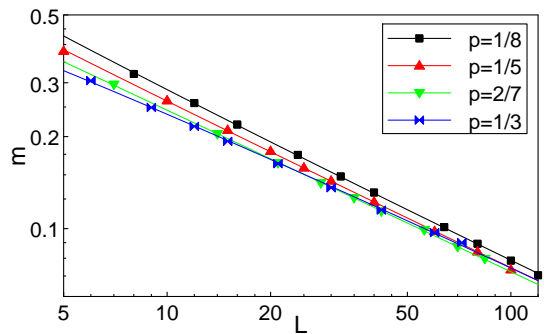}
\caption{ $m$ vs. $L$ at $T_c$ for optimally shaped samples at dilutions $p = \frac 1 8$,
$\frac 1 5$, $\frac 2 7$ and $\frac 1 3$. The statistical error of the data is well below
a symbol size. Solid lines: Fit to $m=aL^{-\beta/\nu}(1+bL^{-\omega_1})$ with
$\beta/\nu=0.53(3)$ and $\omega_1=0.48$.  } \label{fig:betanufit}
\end{figure}

We now turn to the exponents $\beta /\nu$ and $\gamma /\nu$. According to eqs.\
(\ref{eq:m_conv}) and (\ref{eq:chi_conv}), they can be obtained from the $L$-dependence
of the magnetization and susceptibility at $T_c$ of the optimally shaped samples ($L_\tau
=L_\tau^{max}$). In Fig.\ \ref{fig:betanufit} we plot the magnetization data for all four
dilutions $p=\frac 1 8, \frac 1 5, \frac 2 7$, and $\frac 1 3$.
The statistical error of $m$ is below 0.5\%. The plots do not show strong curvature, and
straight power-law fits give values between 0.495 and 0.568 for the exponent $\beta/\nu$.
This could be taken as an indication for nonuniversal behavior. However, given that $z$
is universal, we have also attempted a combined fit of all four curves to
$m(L)=cL^{-\beta/\nu}(1+dL^{-\omega})$ with universal $\beta/\nu$ and $\omega$ (fixed to
the value $\omega_1=0.48$\footnote{Because the $m(L)$ curves do not deviate strongly from
power-law behavior, letting $\omega$ vary does not lead to a stable fit.}) but
dilution-dependent $c$ and $d$. The combined fit works well ($\chi^2 \approx 1.6$) and
gives an asymptotic exponent of $\beta/\nu=0.53(3)$. For comparison, a straight power-law
fit for $p=\frac 2 7$ (which is the dilution where the corrections to scaling
approximately vanish, see above) gives $\beta/\nu=0.527$ in good agreement with the value
from the global fit. We thus conclude that the data display no indication of a
nonuniversal $\beta/\nu$.

\begin{figure}
\includegraphics[width=0.85\columnwidth]{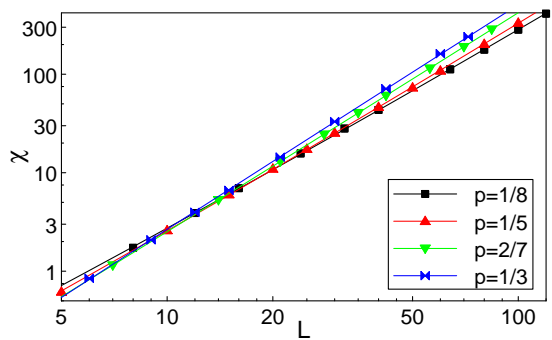}
\caption{ $\chi$ vs. $L$ at $T_c$ for optimally shaped samples at dilutions $p = \frac 1
8$, $\frac 1 5$, $\frac 2 7$ and $\frac 1 3$. The statistical error of the data is well
below a symbol size. Solid lines: Fit to $\chi=aL^{\gamma/\nu}(1+bL^{-\omega_1})$ with
$\gamma/\nu=2.26(6)$ and $\omega_1=0.48$. } \label{fig:gammanufit}
\end{figure}

The analogous plot for the susceptibility data is shown in Fig.\ \ref{fig:gammanufit}.
Here, the statistical error of the data is below 1\%.  Straight power law fits give
effective values between 2.02 and 2.28 for $\gamma/\nu$. The combined fit of all four
curves to the ansatz $\chi(L)=eL^{\gamma/\nu}(1+fL^{-\omega})$ with $\omega=0.48$ gives a
universal asymptotic exponent $\gamma/\nu=2.26(6)$. Again, a straight power-law fit to
the data for $p=\frac 2 7$ gives a value (viz., $\gamma/\nu=2.22$) in good agreement with
the global fit.

The exponents $\beta/\nu$, $\gamma/\nu$ and $z$ are not all independent from each other;
they must fulfill the hyperscaling relation $2\beta/\nu+\gamma/\nu=d+z$. Our values
$\beta/\nu=0.53(3)$, $\gamma/\nu=2.26(6)$, and $z=1.310(6)$ fulfill the hyperscaling
relation within the error bars, indicating that they can indeed be asymptotic values
rather than effective exponents.

\begin{figure}
\includegraphics[width=\columnwidth]{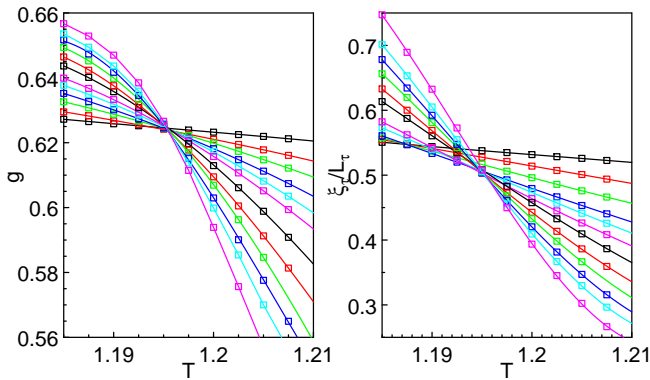}
\caption{Binder ratio $g_{av}$ (left) and $\xi_\tau^{dis}/L_\tau$ (right) as functions of
    temperature $T$ for dilution $p=0.2$. System sizes range from $L=5$ to $L=100$
    with increasing slope.}
\label{fig:g_xi_T_gen}
\end{figure}

After having discussed the simulations right at $T_c$, we now  vary the temperature over
a range in the vicinity of $T_c$, but we consider only samples of ``optimal shape''
$L_\tau=L_\tau^{max}$ to keep the second argument of the scaling functions in eqs.\
(\ref{eq:m_conv}), (\ref{eq:chi_conv}), (\ref{eq:g_conv}), (\ref{eq:xiL_conv}), and
(\ref{eq:xitL_conv}) constant. Fig.\ \ref{fig:g_xi_T_gen} shows the Binder ratio $g_{av}$
and the ratio $\xi_\tau^{dis}/L_\tau$ as functions of temperature for for various $L=5$
to 100 and dilution $p=\frac 1 5$. (The correlation lengths have been calculated in the
usual way via the lowest Fourier components of the spin-spin correlation
function.\cite{CooperFreedmanPreston82,Kim93})
The Binder ratio shows a near perfect crossing point, i.e., the corrections to scaling
for this quantity are very small. In contrast, $\xi_\tau^{dis}/L_\tau$ displays larger
corrections to scaling indicated by the drift of the the crossing point between different
$\xi_\tau^{dis}/L_\tau$ curves with $L$. The behavior of $\xi^{dis}/L$ (not shown) is
very similar to that of $\xi_\tau^{dis}/L_\tau$.

\begin{figure}
\includegraphics[width=0.85\columnwidth]{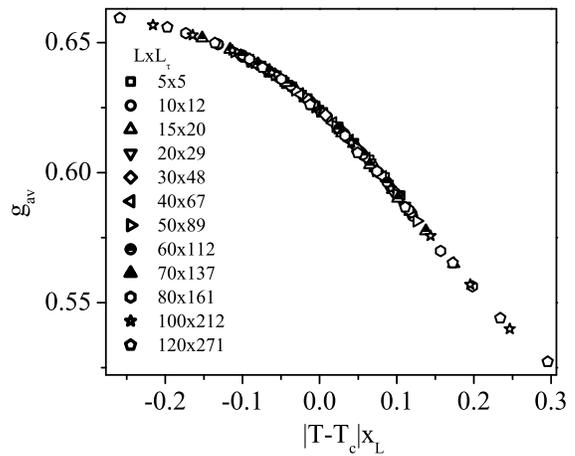}
\caption{Scaling plot of $g_{av}$ vs. $(T-T_c)x_L$ for $p=0.2$.
   $x_L$ is the factor necessary to scale the data onto a master curve.
   The statistical errors of $g_{av}$ are well below a symbol size.}
\label{fig:nuscaling}
\end{figure}

\begin{figure}
\includegraphics[width=0.85\columnwidth]{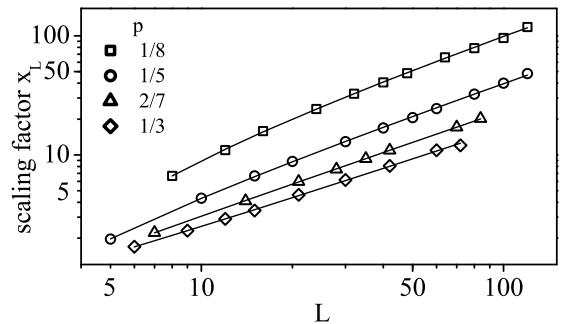}
\caption{Scaling factor vs. $L$ for four disorder concentrations $p = \frac 1 8$, $\frac
1 5$, $\frac 2 7$ and $\frac 1 3$. The statistical errors of $x_L$ are smaller then a
symbol size. Solid lines: Fit to $x_L=gL^{1/\nu}(1+hL^{-\omega_2})$ with $\nu=1.16(3)$
and $\omega_2=0.5(1)$.  } \label{fig:nufit}
\end{figure}

We have therefore used a scaling analysis of the Binder cumulant as our main tool for
determining $\nu$. Fig.\ \ref{fig:nuscaling} shows a scaling plot of $g_{av}$ vs.\
$(T-T_c) x_L$ for impurity concentration $p = \frac 1 5$. (Here $x_L$ is the scaling
factor necessary to collapse the data onto a master curve.)
The quality of the scaling is very good, comparable to that in Fig.\ \ref{fig:zscaling}.
However, since the scaling function lacks the characteristic maximum, the error of the
resulting scaling factor $x_L$ is somewhat larger (1 to 2\%) than that of $L_\tau^{\rm
max}$. The data for other dilutions $p=\frac 1 8, \frac 2 7$ and $\frac 1 3$ lead to
analogous scaling plots. To determine the correlation length exponent $\nu$, we plot the
scaling factor $x_L$ vs.\ $L$ for all four dilutions in Fig.~\ref{fig:nufit}.
Similar to the $L_\tau^{max}$ vs.\ $L$ curves in Fig.\ \ref{fig:zscaling}, the $x_L$ vs.\
$L$ curves show significant curvatures necessitating an ansatz
$x_L=cL^{1/\nu}(1+dL^{-\omega_2})$ that includes corrections to scaling. A combined fit
of all four curves to this ansatz (with universal $\nu$ and $\omega_2$) gives
$\nu=1.16(3)$ and $\omega_2=0.5(1)$. As above, the fit is robust and of high quality
($\chi^2 \approx 1.2$). Importantly, as expected for the true asymptotic exponent, $\nu$
fulfills the inequality $\nu>2/d_\bot$=1.\cite{CCFS86} Note that both irrelevant
exponents $\omega_1$ and $\omega_2$ agree within their error bars, suggesting that the
same irrelevant operator controls the leading corrections to scaling for both $z$ and
$\nu$. For comparison, we have also performed a straight power-law fit for the dilution
where the corrections to scaling approximately vanish, $p=\frac 2 7$. It gives
$\nu=1.12(3)$ in agreement with the value from the global fit.

We have also performed an analogous scaling analysis of $\xi_\tau^{dis}/L_\tau$. Because
of the larger corrections to scaling, the errors of the scale factors (i.e., slopes of
the curves) are significantly higher, but within the error bar, the value of $\nu$ agrees
with that determined from $g_{av}$.

\subsection{Percolation transition at $p=p_c$}
\label{subsec:percolation}

After having discussed the generic dirty quantum rotor phase transition realized for
$p<p_c$, we now turn to the percolation-type transition occurring for $p=p_c$ and
$T<T^\ast$. At this transition, the dynamical fluctuations of the rotors are noncritical;
and the critical behavior is due to the critical geometry of the percolating lattice.
Vojta and Schmalian \cite{VojtaSchmalian05b} have developed a complete scaling theory for
this percolation quantum phase transition. They have also calculated exact exponent
values for the case of two space dimensions, viz., $z=91/48$, $\nu=4/3$, $\beta=5/36$,
and $\gamma=59/12$.

In this subsection, we test these theoretical predictions by performing simulations at
$p=p_c=0.407253$ and $T=0.5$. For two reasons, these calculations require significantly
higher numerical effort than those in the last subsection: (i) Due to the large value of
the dynamical exponent $z$, the ``optimal'' linear size $L_\tau^{max}$ in the time-like
direction increases very rapidly with $L$. For our largest $L=80$, the optimal $L_\tau$
turns out to be $L_\tau^{max}=2030$ leading to a system of 13 million spins. (ii) The
very strong geometric fluctuations of the lattice at the percolation threshold lead to
noisier data. Thus a larger number of disorder realizations has to be averaged.

\begin{figure}
\includegraphics[width=0.85\columnwidth]{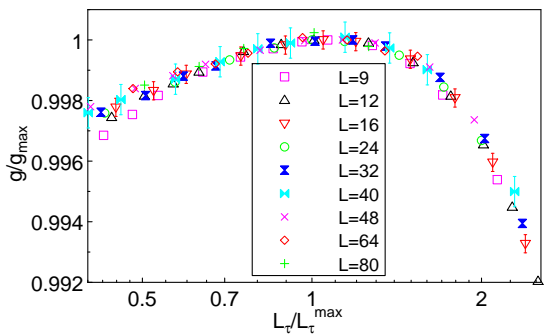}
\caption{ Power-law scaling plot $g_{av}/g_{av}^{max}$ vs. $L_\tau/L_\tau^{max}$ for
$p=p_c=0.407253$ and $T=0.5$. Statistical errors range from $3*10^{-4}$ for the small
sizes to about $6*10^{-4}$ for the larger systems, as indicated. }
\label{fig:zscaling_perc}
\end{figure}

\begin{figure}
\includegraphics[width=0.85\columnwidth]{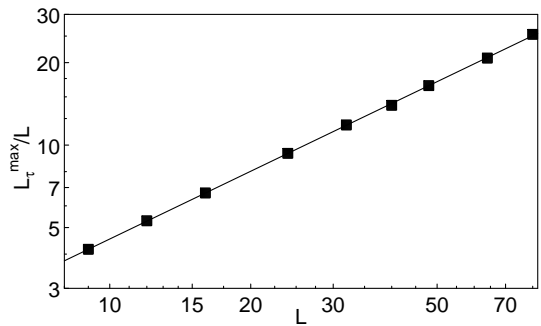}
\caption{ $L_\tau^{max}/L$ vs. $L$ for dilution $p=p_c=0.407253$ and $T=0.5$.  The
statistical error of the data is about a symbol size. Solid line: Power-law fit giving
$z=1.83(3)$.} \label{fig:zfit_perc}
\end{figure}

Figure \ref{fig:zscaling_perc} shows the resulting scaling plot for the Binder cumulant
as a function of $L_\tau/L_\tau^{max}$ for systems of sizes $L=9$ to 80.
The data scale reasonably well, but the quality is clearly less than that of the
corresponding plot for the generic transition (Fig.\ \ref{fig:zscaling}). Moreover there
seems to be a small systematic broadening of the domes with increasing $L$ which is
likely caused by finite-size corrections to the critical lattice percolation problem. The
resulting values of $L_\tau^{max}$ have statistical errors of about 5\%. To determine the
dynamical exponent $z$, we plot $L_\tau^{max}$ vs.\ $L$ in Fig.\ \ref{fig:zfit_perc}.
The data can be well fitted by a power law giving an exponent of $z=1.83(3)$. The
remaining small difference to the theoretical value $91/48 \approx 1.89$ can probably be
attributed to corrections scaling for our rather small $L$. Indeed, a fit of the
$L_\tau^{max}$ vs.\ $L$ data to the ansatz $L_\tau^{max}(L)=aL^{91/48}(1+bL^{-\omega})$
is almost indistinguishable from the power-law fit.

In addition to the Binder ratio we have also analyzed magnetization $m$ and
susceptibility $\chi$ for the optimally shaped samples ($L_\tau=L_\tau^{max}$). In
analogy to Figs.\ \ref{fig:betanufit} and \ref{fig:gammanufit}, the $L$-dependencies of
$m$ and $\chi$ give the exponents $\beta/\nu$ and $\gamma/\nu$, respectively. The $m$
vs.\ $L$ curve shows noticeable upward curvature; and while a power-law fit gives
$\beta/\nu=0.15(3)$, using the ansatz $m(L)=cL^{5/48}(1+dL^{-\omega})$ actually leads to
a significantly better fit (lower $\chi^2$). For the susceptibility, a power-law fit of
the $\chi$ vs. $L$ data gives $\gamma/\nu=3.51(5)$ which is somewhat smaller than the
theoretical value of $59/16 \approx 3.68$. However, a fit to the ansatz
$\chi(L)=eL^{59/16}(1+fL^{-\omega})$ is of comparable quality.

\subsection{Multicritical point}
\label{subsec:multicritical}

In this last subsection on results, we consider the multicritical point $(p_c,T^\ast)$
that separates the generic transition from the percolation transition. In contrast to the
percolation transition, no quantitative theoretical predictions are available for the
multicritical point. We have performed simulations at $p=p_c=0.407253$ and
$T=T^\ast=0.791$ for $L=9$ to 64. Because of the simultaneous presence of critical
geometric and dynamical fluctuations, the necessary number of disorder realizations is
even larger than for the percolation transition. We have used between 10000 and 50000
realizations, depending on system size.

\begin{figure}
\includegraphics[width=0.85\columnwidth]{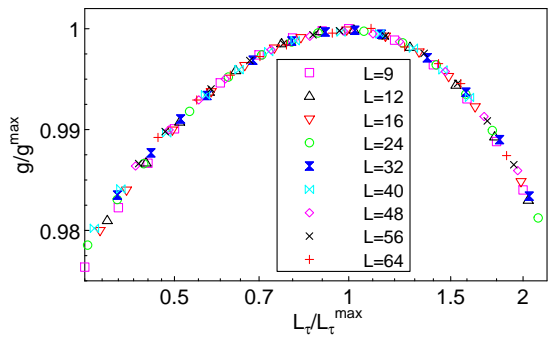}
\caption{ Power-law scaling plot $g_{av}/g_{av}^{max}$ vs. $L_\tau/L_\tau^{max}$ for the
multicritical point at $p=p_c=0.407253$ and $T=T^\ast=0.791$. The statistical error of
the data is about a symbol size.} \label{fig:zscaling_mc}
\end{figure}

\begin{figure}
\includegraphics[width=0.85\columnwidth]{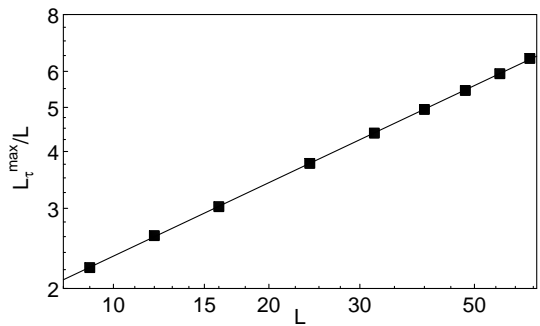}
\caption{ $L_\tau^{max}/L$ vs. $L$ for the multicritical point $p=p_c=0.407253$ and
$T=T^\ast=0.791$. The statistical error is well below a symbol size. Solid line:
Power-law fit giving $z=1.54(2)$.} \label{fig:zfit_mc}
\end{figure}

Figure \ref{fig:zscaling_mc} shows the resulting scaling plot for the Binder cumulant as
a function of $L_\tau/L_\tau^{max}$.
The data scale very well, giving statistical errors for $L_\tau^{max}$ of approximately
2\%. Note that the data show a very slight systematic broadening of the domes with
increasing $L$. It is much weaker than for the percolation transition at $T=0.50$ (Fig.\
\ref{fig:zscaling_perc}), but probably, it can also be attributed to finite-size
corrections to the critical lattice percolation problem. Fig.\ \ref{fig:zfit_mc} shows a
log-log plot of $L_\tau^{max}$ vs $L$.
The curve does not show a discernable deviation from a straight line, and a power-law fit
gives $z=1.54(2)$. Given the fact that the same analysis gave a very slightly too small
$z$-value at the percolation transition, the true asymptotic exponent may be a few
percent higher than the fit result. We have also determined the exponents $\beta/\nu$ and
$\gamma/\nu$ from the $L$-dependencies of the magnetization and susceptibility for the
optimally shaped samples. As at the percolation transition, the $m(L)$ curve shows some
upward curvature; and while a power-law fit gives $\beta/\nu=0.40(3)$, the true
asymptotic exponent may be a bit lower. Unfortunately, our $L$-range is not wide enough
for a stable fit to an ansatz that includes corrections to scaling (and thus has two
unknown exponents and two prefactors). In contrast, the $\chi(L)$ curve does not show any
deviations from power-law behavior, and a fit gives $\gamma/\nu=2.71(3)$. Again, from the
analogy with the percolation transition, the true asymptotic exponent may be slightly
higher.

\section{Conclusions}
\label{sec:conclusions}

We have investigated the quantum phase transitions of a two-dimensional site-diluted O(3)
rotor model by performing large-scale Monte-Carlo simulations of the equivalent classical
model, a three-dimensional classical Heisenberg model with linear defects. In this final
section we summarize the results and relate them to a recent classification
\cite{VojtaSchmalian05,Vojta06} of phase transitions with quenched disorder. We also
compare our findings with previous work on this and related problems; and we consider
experiments.

The two-dimensional site-diluted O(3) rotor model has two quantum phase transitions, (i)
a generic transition for dilutions $p$ below the percolation threshold $p_c$ of the
lattice and (ii) a quantum percolation transition at $p_c$. These transitions are
separated by a multicritical point. Our calculations have shown that the critical
behavior of all these transitions is of conventional power-law type. In contrast, the
Ising version of our model, the diluted 2d random transverse-field Ising model, shows an
infinite-randomness critical point.\cite{SenthilSachdev96,PYRK98,MMHF00} To study the
generic transition, we have considered four different dilutions. Combined fits of all
data sets have allowed us to systematically include corrections to scaling. In this way,
we have provided strong evidence that the critical behavior of the generic transition is
universal, i.e., independent of the dilution. Because of the high numerical effort, we
have considered only one data set for the percolation transition. This one data set does
not allow us to extract the leading exponents \emph{and} the corrections to scaling from
a free fit of the numerical data. However, including corrections to scaling when fitting
the data to the theory of Ref.\ \onlinecite{VojtaSchmalian05b} leads to a very good
agreement. For the multicritical point, there is only one data set, and no quantitative
theoretical results exist. Therefore, our results for the multicritical behavior are on
somewhat less firm ground because they do not contain corrections to scaling. Our
exponent values are summarized in table \ref{table:exponents}.
\begin{table}
\begin{tabular}{ccccc}
\hline
Exponent    & Generic  &  Multicritical  & Percolation & Perc. (theory)\\
\hline
$z$         & 1.310(6) &  1.54(2)        & 1.83(3)     & 91/48   \\
$\beta/\nu$ & 0.53(3)  &  0.40(3)        & 0.15(3)     &  5/48   \\
$\gamma/\nu$& 2.26(6)  &  2.71(3)        & 3.51(5)     & 59/16   \\
$\nu$       & 1.16(3)  &                 &             &         \\
\hline
\end{tabular}
\caption{Numerical results for the critical exponents. For the generic transition,
the values are from the combined fit of all four datasets for dilutions $1/8$, $1/5$,
$2/7$, and $1/3$. For the multicritical point and the percolation transition the values
are from straight power-law fits. The numbers in brackets give the \emph{statistical} error
of the last given digits. For the percolation transition we also give the theoretical
results from Ref. \onlinecite{VojtaSchmalian05b}. The numerical data are compatible with
these values if one allows for corrections to scaling.}
\label{table:exponents}
\end{table}

Recently, a general classification has been suggested for phase transitions with weak
(random-$T_c$ type) quenched disorder and short-range interactions.
\cite{VojtaSchmalian05,Vojta06} According to this classification, the type of critical
behavior depends on the effective dimensionality $d_{RR}$ of the defects or,
equivalently, the rare regions. Three cases can be distinguished. (A) If $d_{RR}$ is
below the lower critical dimension $d_c^-$ of the problem, rare region effects are
exponentially small. As a result, the transition is sharp, and the critical point is of
conventional power-law type. (B) In the second class, with $d_{RR}=d_c^-$, rare regions
are much more important. A sharp transition still exists, but the critical point is
controlled by an infinite-randomness fixed point with activated scaling. In addition,
there are strong power-law Griffiths effects. (C) Finally, for  $d_{RR}>d_c^-$, the rare
regions can order independently leading to a destruction of the sharp phase transition by
smearing.

In the problem considered here, $d_{RR}=1$ because the defects are linear. The lower
critical dimension of the Heisenberg universality class is $d_c^-=2$. Therefore,
$d_{RR}<d_c^-$, and our model should be in class A with conventional power-law critical
behavior. Our numerical results are thus in complete agreement with the above general
rare-region based classification scheme.

Let us compare our results to previous work. The qualitative structure of the phase
diagram, viz., the fact that long-range order survives for all dilutions up to and
including the percolation threshold agrees with earlier quantum Monte-Carlo simulations
for the bilayer quantum Heisenberg antiferromagnet \cite{Sandvik02,VajkGreven02} and with
analytical results for diluted magnets \cite{MuccioloCastroNetoChamon04} as well as O(2)
rotors.\cite{BMSV06} Sandvik\cite{Sandvik02} and Vajk and Greven\cite{VajkGreven02}
studied the multicritical point at $p=p_c$. They found a dynamical exponent of
$z\approx1.3$ significantly lower than our result of 1.54. More recently,
Sandvik\cite{Sandvik06} reported a somewhat larger value of $z = 1.36$, but it is still
well below our result. The reasons for this discrepancy are presently not fully
understood. Possible explanations include a failure of the quantum to classical mapping
(which we consider unlikely) and corrections to scaling of the lattice percolation
problem. In this context it is worth noting that the scaling properties of the lattice
enter our calculations in a different way than that of Ref.\ \onlinecite{Sandvik06}. Our
analysis of the Binder ratio works with linear extensions in space an time directions,
directly giving $z$. In contrast, Ref.\ \onlinecite{Sandvik06} analyzes the temperature
dependence of the susceptibility and effectively measures $D_f/z$ with $D_f$ being the
fractal dimension of the percolation cluster. It is clear that corrections to scaling, if
any, will enter the two calculations very differently.

Vajk and Greven \cite{VajkGreven02} also quoted exponents for $p<p_c$. At dilution
$p=0.25$ they find $z=1.07$ and $\nu=0.89$, different from our results. However, as the
authors of Ref.\ \onlinecite{VajkGreven02} pointed out, a value of $\nu<1$ violates the
inequality $\nu>2/d$, indicating that it represents an effective rather than an
asymptotic exponent. It would also be useful to compare our exponents with analytical
results. To the best of our knowledge, the only quantitative result for the generic
transition is a resummation of the 2-loop $\epsilon$-expansion.
\cite{BlavatskaFerberHolovatch03} The predicted exponents significantly differ from ours;
but they also violate the inequality $\nu>2/d$, casting doubt on their validity.

While no quantitative analytical results exist for the multicritical point, there is a
complete scaling theory for the percolation transition in the diluted rotor model,
\cite{VojtaSchmalian05b} and the exponents in two dimensions are known exactly. Our
numerical data are in excellent agreement with the exact values if one allows for
corrections to scaling. Even if corrections to scaling are not included, the differences
between the theoretical and numerical exponent values are only a few percent. Originally,
this scaling theory was thought to apply not only to the rotor model but also to a
site-diluted Heisenberg antiferromagnet; and there is some numerical
evidence\cite{YuRoscildeHaas05} in support of the relation $z=D_f$ predicted by the
scaling theory.  However, a recent exact diagonalization and quantum Monte-Carlo study
\cite{WangSandvik06} finds a dynamical exponent $z\approx 2 D_f$ for the site-diluted
Heisenberg antiferromagnet but $z \approx D_f$ for the dimer-diluted bilayer (which is
equivalent to a rotor model).

 Finally, we discuss experiments. Chemical
doping, i.e., random replacement of magnetic by non-magnetic ions, e.g., Cu by Zn in
YBa$_2$Cu$_3$O$_6$, in both single-layer and bilayer antiferromagnets realizes site
rather than dimer dilution. As discussed at the end of section \ref{subsec:model}, this
leads to random Berry phases and a completely different physical picture. The most
promising way to achieve dimer dilution is the introduction of strong antiferromagnetic
intra-dimer bonds at random locations. Thus we propose to study magnetic transitions in
bond-disordered systems; those transitions can be expected to be in the same universality
class as the one studied here. One candidate material -- albeit 3d -- is (Tl,K)CuCl$_3$
\cite{OosawaTanaka02} under pressure; interesting quasi-2d compounds are
SrCu$_2$(BO$_3$)$_2$ or BaCuSi$_2$O$_6$, where suitable dopants remain to be found. Our
results may also be interesting for some single-layer Zn doped cuprate antiferromagnets
that have been speculated to be parametrically close to the multicritical point of the
rotor model. \cite{Sandvik02,VajkGreven02} Moreover, the qualitative properties of the
phase diagram and the critical behavior are also important for disordered Josephson
junction arrays or diluted bosons in optical lattices.

\section*{Acknowledgements}
We thank Matthias Vojta for a collaboration in the early stages of this work. We have
also benefitted from discussions with Stephan Haas, Martin Greven, Anders Sandvik, and
Joerg Schmalian. This work has been supported in part by the NSF under grant nos.
DMR-0339147 and PHY99-07949, by Research Corporation, by NSERC of Canada, and by the
University of Missouri Research Board. We are also grateful for the hospitality of the
Aspen Center for Physics and the Kavli Institute for Theoretical Physics, Santa Barbara
where parts of this work have been performed.


\bibliographystyle{apsrev}
\bibliography{../00Bibtex/rareregions}

\end{document}